\documentclass[12pt]{article}
\usepackage{graphics}
\usepackage{graphicx}
\usepackage{epsfig}
\usepackage{wrapfig}

\begin{document}

\sloppy
\begin{flushright}{SIT-HEP/TM-17}
\end{flushright}
\vskip 1.5 truecm
\centerline{\large{\bf Incidental Brane Defects }}
\vskip .75 truecm
\centerline{\bf Tomohiro Matsuda
\footnote{matsuda@sit.ac.jp}}
\vskip .4 truecm
\centerline {\it Laboratory of Physics, Saitama Institute of
 Technology,}
\centerline {\it Fusaiji, Okabe-machi, Saitama 369-0293, 
Japan}
\vskip 1. truecm
\makeatletter
\@addtoreset{equation}{section}
\def\theequation{\thesection.\arabic{equation}}
\makeatother
\vskip 1. truecm

\begin{abstract}
\hspace*{\parindent}
In the models of brane construction, the isometry of a compactified
 space might be broken by branes.
In four-dimensional effective Lagrangian, the breaking of the isometry is
 seen as the spontaneous breaking of the corresponding effective symmetry.
Then it seems natural to 
expect that there are various kinds of defects that will be implemented
by the spontaneous symmetry breaking.
These defects are parametrized by the brane positions.
In this paper we consider two kinds of such ``brane defects'', which are
formed by the local fluctuations of the locations of branes along their 
transversal directions.
The fluctuation of a brane position might leads to winding (or wraping)
 around a non-contractible circle of the compactified space.
These ``primary'' brane defects are already discussed by several authors.
On the other hand, if there are multiple branes in the compactified
 space and their configuration in a compactified
 space is determined by  
the potential that depends only on their relative positions,
one might find incidental symmetry in the effective potential,
which is spontaneously broken by branes.
We examined the latter ``incidental'' symmetry breakings and stable
 defect configurations.
We paid special attention to the difference between ``primary'' brane defects. 
\end{abstract}

\newpage
\section{Introduction}
\hspace*{\parindent}
In spite of the great success of quantum field theory and classical
Einstein gravity, there is still no consistent unification scenario in which
quantum gravity is successfully included.
Perhaps the most promising scenario in this direction is string
theory, in which consistency of the quantum gravity is ensured by a
requirement of additional dimensions.
Originally the size of extra dimensions was assumed to be as small
as $M_p^{-1}$. 
However, later observations showed that there is no reason
to require such a tiny compactification radius\cite{Extra_1}.
In this respect, what we had seen in the old string theory was a tiny 
part of the whole story.
In the new scenario, the compactification radius (or the fundamental
scale) is an unknown parameter that should be determined by
observations.  
In models with large extra dimensions, 
the observed Planck mass is obtained by the relation $M_p^2=M^{n+2}_{*}V_n$,
where $M_{*}$ and $V_n$ denote the fundamental scale of gravity
and the volume of the $n$-dimensional compact space.
In this scenario the standard model fields are expected to be localized
on a wall-like structure and the graviton propagates in the bulk.
The most natural embedding of this picture in the string theory context 
is realized by a brane construction.
Thus it is quite important to construct the models of the brane world 
where the observed spectrum of the standard model is included
in the low energy effective theory.
Moreover, we know that sometimes the cosmology of the models for the
braneworld (or models with large extra dimensions) seems quite
peculiar.\footnote{Constructing successful models for  
inflation with a low fundamental scale is still an interesting
problem\cite{low_inflation, matsuda_nontach}.
Baryogenesis and inflation in models with a low fundamental scale are
discussed in \cite{low_baryo, low_AD, ADafterThermal}.
We think constructing models of particle cosmology with large
extra dimensions is very important since we are expecting that future
cosmological 
observations would determine the fundamental
scale of the underlying theory.}
We know historically that the characteristic features of
phenomonological models are revealed by discussing their cosmological
evolutions. 

In conventional models with more than four dimensions, compactified
space has a topology $K$.  
The isometry of $K$ is seen as a gauge symmetries of the effective
four-dimensional theory.
However, in theories where our world is a domain wall (or brane)
embedded in the extra space, one might expect that the isometry is
broken by the existence of such object.
\footnote{If the compactified space is warped by branes,
effective symmetry is explicitly broken in low-energy Lagrangian.
In this case the masses of the pseudo-Nambu-Goldstone (pNG) bosons are
determined by the warp factor\cite{NG-iso-DabadoMaroto}.}
Even if an isometry of the compactified space remains as an exact symmetry in
the effective Lagrangian, it might be broken spontaneously by the
existence of a brane. 
A brane localized in $K$ breaks spontaneously the isometries of $K$,
which is seen as a Higgs effect in the four-dimensional effective
theory\cite{NG-iso-DabadoMaroto}.
As a brane localized in $K$ breaks the symmetry
of the effective four-dimensional theory, one might expect that there
exists a topologically nontrivial mapping of the brane position,
which corresponds to strings or monopoles in conventional field theory.
Of course, topological defects of similar mappings can be
constructed within classical Kaluza-Klein setup without branes, such as 
Kaluza-Klein strings or monopoles.
However, our main concern in this paper is to examine defects that are defined
by a nontrivial mapping of the brane positions in extra dimensions.
In this paper, we call them ``brane defects''.

In generic cases, brane defect is constructed from a non-contractible
 winding (or wrapping) of a brane position around extra dimension(s).
In this case, the parameter of the mapping is defined by absolute
 coordinate in the compactified space.
In this paper, we call them ``primary'' brane defects.
Primary brane defects are discussed by several authors.
Strings (including Alice strings) and monopoles are discussed in
 ref.\cite{Dvali_alice}, and skyrmions are discussed in
 ref.\cite{Skyrmion}.
At first sight, one might feel some doubt about these configurations.
Recalling a typical string configuration in conventional field theory, 
one will find that the broken symmetry is restored
 in the core.
On the other hand, in brane defects that utilize the brane position in
compactified space, one cannot simply expect such restoration in the core.
Is there any possible mappings that might not be singular in the core?
An answer to this question is given by Dvali et al.\cite{Dvali_alice}.
The authors constructed branes from the kink configurations of a scalar
 field and showed that the kink (brane) is delocalized in the core, which 
solves the naive singularity.
These ``primary'' defects have many interesting properties.
%%deleted sentence
However, in generic models, a secondary weak inflation is sometimes
required, which dilutes  
existing particles and defects.
After the dilution of cosmological defects, it is hard for the defects to
influence the later cosmology.
In this case, some peculiar mechanisms will be required for the defects to be 
produced after secondary weak inflation.
Thus, we think it is quite interesting if one could find brane defects
that have the same (or similar) peculiar properties but are safely
produced after weak inflation.\footnote{In this case one should be careful
about cosmological constraints, because cosmological defects sometimes
put serious constraint on the model, if they are produced\cite{vilenkin,
 matsuda_wall}.}

In this paper, we focus our attention to the {\bf relative} coordinates of
branes.
Our idea is quite simple.
In general, relative coordinates might have
{\bf incidental} symmetries, which could be used to construct brane defects.  
When branes change their relative positions along extra dimensions as a
 consequence of local fluctuations, the effective symmetry of the
 relative brane-brane coordinate might allow stable defect formation in 
 four-dimensional spacetime.  
We will show some explicit examples of these incidental brane defects.
We then 
show that they are produced after thermal brane inflation\cite{Thermal}.
For the simplest example, let us consider two branes located in large
$S^1$ that is perpendicular to the branes.
Here we assume that the
distance between the two branes is fixed by the mechanism that is discussed 
in ref.\cite{Thermal}. 
In this case, it is quite easy to see that 
degenerated vacua are formed when two branes exchange their positions
along the extra dimension.
If the exchange happens in a local domain of 
four-dimensional spacetime, a domain wall will appear on the surface of
the domain, which interpolates between two degenerated vacua.
This naive idea seems successful, which will be discussed in section 2
by using an explicit kink(brane) configuration of a scalar field.
Now let us imagine that there are two large extra dimensions that are
perpendicular to the branes.
If the two branes wind around each other in the compactified 
space, a peculiar type of string will be formed in the four-dimensional
 spacetime. 
Along this line of thought, it is possible to construct monopole-like
configurations when there are {\bf three} extra dimensions.
More peculiar example would be a time-dependent configuration that 
will correspond
to the so-called Q-balls in conventional four-dimensional theory.

In general, the cosmological production of incidental brane
defect will be easier than primary brane defects.\footnote{In
ref.\cite{Tachyon-defect-production}, brane inflation with 
brane-brane interactions at an angle is analyzed.
When branes collide, tachyon condensation appears and it generically
allows the formation of lower-dimensional branes.
After brane inflation, when the ground state string mode becomes
tachyonic, the typical particle horizon size is larger than the sizes of
the compactified dimensions, which means that the universe is
homogeneous in the compactified dimensions.
Thus, one might think that the Kibble mechanism does not produce
defects in the compactified directions\cite{Tachyon-defect-production}.
The resulting lower-dimensional defects are
$D(p-2)$-branes wrapping the 
same compactified cycles as the original $p$-branes, with $(3-2)$
uncompactified dimension, which turn out to be the cosmological strings.
Thus, they have concluded that other defects, such as domain walls or
monopoles, are not produced after brane inflation. 
These statements are normally true, but there are some exceptions.
The Kibble
mechanism might take place in the uncompactified directions
for the fluctuations of brane positions along extra dimensions. 
The productions of certain types of domain walls, monopoles and other
defects are possible just after (thermal) brane inflation.}
Some peculiar cosmological implications of these incidental brane defects
are discussed in the forthcoming papers\cite{matsuda_prepare}.

\section{Domain walls}
\hspace*{\parindent}
In this section we consider two types of domain wall configuration,
which are produced by the variations of the relative brane positions.
In the first example we consider two branes constructed from an explicit kink
configuration of a scalar field, while in the second example
we consider three {\bf intersecting} branes.

Our first example\footnote{See fig.1} is a domain wall in the
four-dimensional spacetime, 
which is produced by exchanging the positions of two domain walls (branes) 
along extra dimension.
The two walls are located at the fixed points of an
orbifold\cite{orbifold_wall}. 
Here we consider the simplest model where a single real
scalar field $\phi$ lives in five dimensions.
The extra dimension $x_5$ is in the interval $[0,2L]$.
Imposing a boundary condition on the field that are periodic up to the
$Z_2$ symmetry of the Lagrangian, the extra dimension becomes an
orbifold.
The Lagrangian is
\begin{equation}
{\cal L}=\frac{1}{2}\partial^{\mu} \phi \partial_{\mu}\phi
-\frac{\lambda}{4}\left(\phi^2-v^2 \right)^2,
\end{equation}
where the coupling $\lambda$ is real and $v^2 >0$.
The Lagrangian is invariant under the transformation
\begin{equation}
\phi(x,x_5)\rightarrow \phi'(x,x_5)\equiv -\phi(x,L-x_5).
\end{equation}
The required boundary condition is 
\begin{eqnarray}
\label{orbifold_boundary}
\phi(x,-x_5) &=&  \phi'(x,L-x_5) = -\phi(x,x_5)\nonumber\\
\phi(x,L+x_5) &=& \phi'(x,x_5)   = -\phi(x,L-x_5).
\end{eqnarray}
Although the boundary condition (\ref{orbifold_boundary}) requires
vanishing scalar field on the fixed points at $x_5=0$ and
$x_5=L$, the scalar field will develop a vacuum expectation value in the
bulk. 
In this case, a kink (anti-kink) configuration appears at the fixed
points.
It is easy to find two degenerated vacua that is produced by
flipping the sign of $<\phi>$. 
When the sign of $<\phi>$ is flipped along a direction of
four-dimensional spacetime, the positions of the kink and the anti-kink
in the fifth dimension will be exchanged.
A domain wall is formed in four-dimensional spacetime,
which interpolates between the domains of $\phi <0$ and $\phi >0$.
In this case the degenerated vacua are formed by exchanging the positions of
kink and anti-kink in the fifth dimension, which is induced by flipping
the sign of $<\phi>$ in the four-dimensional spacetime.\footnote{
To make domain walls cosmologically harmless, the energy difference
$\epsilon$ between two quasi-degenerated vacua must satisfy the condition,
\cite{vilenkin}
\begin{equation}
\epsilon\geq\frac{\sigma^{2}}{M_{p}^{2}},
\end{equation}
where $\sigma$ is the tension of the domain wall.
The degeneracy is broken if there is another scalar field $\phi_2$
that satisfies the same boundary condition.
Although the $Z_2^{sim}$ symmetry that corresponds to the {\bf
simultaneous} flips of $<\phi>$ and $<\phi_2>$ will remain, the $Z_2$
symmetries of their {\bf independent} flips are explicitly broken
if there is an effective cross term
$\sim (\phi\phi_2)^{2n+1}$, where $n$ is integer.
In this case, the role of $\phi_2$ corresponds to a constant that is called
``odd mass'' in ref.\cite{orbifold_epsilon}.}

Now let us consider what happens in the core of the
incidental brane defect.
In the core of the defect, the vacuum expectation value of the
scalar field $<\phi>$ vanishes.
In this case, it will be natural to think that the walls(branes) on
which we are living are delocalized in the core.
As is discussed in ref.\cite{Dvali_alice} for primary brane defects, 
the localized matter fields feel fifth dimension in the core.
Thus, it acts as  ``holes'' or ``windows'' to the extra 
dimensions.\footnote{This point is discussed in \cite{Dvali_alice}
for primary brane defects.
However, domain walls are not discussed in previous papers.}

Even if we do not invoke an explicit construction of branes as defects from 
scalar fields, incidental domain walls might appear in any models
where multiple branes live in one perpendicular dimension.
Let us assume that in a local domain of four-dimensional spacetime
 branes exchange their positions along fifth dimension.
If the new vacuum is (at least locally) stable, it is possible to 
construct domain walls interpolating between two domains in the
four-dimensional spacetime.
In this case, the exchanged branes will overlap in the core of the 
domain wall.\footnote{In the explicit brane construction from scalar
fields, there are at least two types of symmetry breaking.
As a brane is constructed by a kink configuration of a scalar
field, a symmetry breaking occurs when the scalar field develops
non-zero vacuum expectation value.
Branes are produced at this time. 
If supersymmetry remains in the bulk or at least softly broken, the
interaction 
between two branes might be well suppressed.
In this case one can expect that thermal effects might stabilize 
two or more branes to coincide at a point\cite{Thermal}.
Then the second symmetry breaking occurs when branes fall apart.
Assuming supersymmetry, the scales of the two phase transitions will be
hierarchically discriminated.}
The ``overlap'' is interesting from a cosmological point of view,
since the suppressed couplings might be enhanced in the
core\cite{low_baryo, ADafterThermal, Dvali_bary}. 

Let us discuss our second example.
Another type of the incidentally degenerated vacuum 
might appear in the models of interesting brane world, where three branes are
intersecting in the compactified dimensions.
Here we consider the simplest example in ref.\cite{Int-brane-world, yukawa}.
What we would like to see is the Yukawa couplings in the quark sector.
The Yukawa coupling among two chiral fermions and one Higgs boson
cannot appear from the perturbative effects of the string theory, but
induced by worldsheet instanton corrections for the corresponding
triangle that has three boundaries of the intersecting branes and three
vertices where matter fields live.

To be more concrete, here we consider the simplest case and derive the
expression for Yukawa couplings.
When computing a sum of worldsheet instantons, the simplest example
comes from D-branes wrapping 1-cycles in a $T^2$, where branes are
intersecting at one angle.
Here we associate each brane to complex number $z_{\alpha},
(\alpha=a,b,c)$, 
\begin{eqnarray}
z_a&=&R\times (n_a+\tau m_a)\times x_a\nonumber\\
z_b&=&R\times (n_b+\tau m_b)\times x_b\nonumber\\
z_c&=&R\times (n_c+\tau m_c)\times x_c.
\end{eqnarray}
Here $(n_{\alpha},m_{\alpha})\in {\bf Z}^2$ denote the 1-cycle the brane
$\alpha$ wraps on $T^2$ and
$x_{\alpha} \in {\bf R}$ is an arbitrary number.
$\tau$ is the complex structure of the torus.
These branes are given by a straight line in ${\bf C}$. 
The triangle corresponding to a Yukawa coupling must involve three branes,
which has the form $(z_a,z_b,z_c)$ with $z_z+z_b+z_c=0$.
The solution is 
\begin{equation}
x_{\alpha}=I_{\beta\gamma}x/d,
\end{equation}
where $x=x_0+l, x_0 \in {\bf R}, l \in {\bf Z}$ and
$d=g.c.d.(I_{ab},I_{bc},I_{ca})$.
Here $I_{\beta\gamma}$ stands for the intersection number of branes
$\beta$ and $\gamma$.
Indexing the intersection points, one can obtain a simple expression for 
$x_0$\footnote{See ref.\cite{yukawa} for more detail.},
\begin{equation}
x_0(i,j,k)=\frac{i}{I_{ab}}+\frac{j}{I_{ca}}+\frac{k}{I_{bc}}
+\frac{I_{ab}\epsilon_c + I_{ca}\epsilon_b +I_{bc}
\epsilon_a}{I_{ab}I_{bc}I_{ca}},
\end{equation}
where the parameter $\epsilon_\alpha$ correspond to shifting the
positions of the three branes.
Using this solution, one can compute the areas of the triangles whose
vertices lie on the triplet of intersections $(i,j,k)$,
\begin{equation}
A_{ijk}(l)=\frac{1}{2}(2\pi)^2 A|I_{ab}I_{bc}I_{ca}|\left(
x_0(i,j,k)+l\right)^2
\end{equation}
where $A$ represents the K\"ahler structure of the torus.
The corresponding Yukawa coupling is given by 
\begin{equation}
\label{yukawa_formula}
Y_{ijk}\sim \sigma_{abc}\sum_{l\in{\bf Z}}
exp\left(-\frac{A_{ijk}(l)}{2\pi\alpha'}\right),
\end{equation}
where $\sigma_{ijk}=sign(I_{ab}I_{bc}I_{ca})$ is a real phase.

Now our question is how one can determine the areas of the triangles.
A perturbative force between branes can
produce potential for the distance between two branes.
However, it is obvious that this force cannot affect the area of a
triangle when branes are {\bf intersecting}.
On the other hand, one can see from eq.(\ref{yukawa_formula}) that
almost all the parameters are determined if the windings of the
branes and the structure of the manifold are fixed by some mechanisms.
The only ambiguity that might remain at low energy effective theory is
one parameter of three $\epsilon_\alpha$, which corresponds to shifting
the relative brane position. 
For the area of a triangle, only one of the three parameters
$\epsilon_\alpha$ is independent.

An effective potential for the area of a triangle is obtained by
considering a well-known 1-loop correction from fermion loops,\cite{Kolb-turner}
\begin{equation}
\label{effective}
\Delta V(\phi_c)=-\frac{3}{64\pi^2} Y_{ijk}^4 \phi_c^4 
ln\left(\frac{\phi_c^2}{\mu^2}\right),
\end{equation}
where $\phi_c$ denotes the classical field.
From eq.(\ref{effective}) and (\ref{yukawa_formula}), one can easily 
see that the 1-loop correction stabilizes the area of the
triangle.\footnote{See fig.2}

Because of the exponential form of the potential, intersecting branes
will be 
stabilized when one of the areas of the three triangles vanishes.
Because of phenomonological requirements and the corresponding brane
setups, three triangles cannot 
shrink simultaneously to a point.
Thus it is possible to construct stabilized models in which one of the
three Yukawa 
couplings becomes large, while others remain (hierarchically)
small.
In general, at least three triangles are included in models for
intersecting braneworld, which correspond to three generations in the
standard model.
Then it leads to (at least) three degenerated vacua where each triangle 
might shrink.
If three generations
are geometrically equal, there is no sensible reason
why the third generation is cosmologically selected to become the heaviest.
These degenerated vacua suggest the existence of incidental domain walls.

Here we stress that our idea in this section is quite generic.
When there are multiple branes in the compactified space, their positions
must be determined by some mechanisms.
The most obvious forces are induced by interactions between two
distant branes, which produce an effective potential that depends only
on the absolute value of the distances.
If there are many branes in the compactified space and their
configuration is determined by the potential of this type,
it is quite natural to expect incidental domain walls that are produced
by the fluctuations of the brane positions along extra dimensions.
The situation is similar to the conventional defect in the crystal. 
In general, as we have discussed above, interactions among more than two
branes are less effective than the perturbative force between two branes. 
As an example, we considered triangle interaction among intersecting three
branes and showed that it is possible to construct 
domain walls in intersecting brane models.

\section{Strings}
\hspace*{\parindent}
The primary brane defects that were discussed in 
\cite{Dvali_alice,Skyrmion} are constructed by a single brane position
along extra dimensions.
On the other hand, in more generic situations, one would expect multiple
branes in a compactified space.
Thus our starting point in this section is to add branes to the well-known
models and examine the difference.
Here we mainly consider a compactified space $S^2$.
If there is only one brane in the compactified space, the symmetry
breaking is $SU(2)\rightarrow U(1)$, which forbids stable string
configuration.
Let us recall that in conventional field theory many models have been
discussed to circumvent this difficulty.
For example, if the symmetry is broken to  $SU(2)\rightarrow U(1)\times
Z_2$, one can find stable string configuration\cite{vilenkin-book}.
This possibility is examined in ref.\cite{Dvali_alice}
by adding an additional brane at the opposite pole of $S^2$.
Identifying two branes, one can obtain $Z_2$ symmetry.

First we consider a looser condition.
For simplicity, we assume large $S^2$ where two branes are located.
Distance between branes is assumed to be stabilized by a
mechanism.
We also assume that the distance between the two branes is stabilized at 
a distance scale smaller than the radius of the compactified 
space.\footnote{A possible mechanism for stabilization is discussed in
ref.\cite{Thermal}.
If the distance is not stabilized but there is only a repulsive force
between them, two branes will repel to north and south poles.
In this case, unwanted $U(1)$ will remain.}
Considering the idea of thermal brane inflation\cite{Thermal},
it is natural to expect that the two
branes might glue together to a point by thermal effects during a period
of the Universe. 
During this period, the symmetry is restored to $U(1)$.
Then the two branes will fall apart, which induces symmetry breaking 
$U(1)\rightarrow I$, because we have assumed that the distance between
branes is smaller than the 
radius of the compactified space.
If one considers a limit where one brane is much heavier than
the other, 
one can easily find an effective description of the ``light'' brane
fluctuation around a probe brane, which becomes quite 
similar to the conventional field theory in the lowest order 
expansion\cite{NG-iso-DabadoMaroto,brane-effective}.
In this case, an incidental string is formed by the position of a light
brane, parametrized by the windings around the heavy brane.
Unlike the primary strings that winds around $S^1$ compactified space,
one do not have to worry about singularity at the string core.
In the core of the incidental brane string, two branes will coincide to
restore the $U(1)$ 
symmetry.
This point is quite different from the primary string that requires
delocalization of the brane in the core to solve the singularity.

To make our discussion more clear, here we consider an explicit example.
As we are considering ``brane'' that might be delocalized in the core,
it will be helpful to consider vortices from a scalar field in $S^2$
compactified space. 
An explicit construction of such vortices is already discussed in 
\cite{Vortex-J}.
Here we do not repeat the details of the calculations.
We simply examine the obtained solution for three vortices.
The model is defined in the spacetime ${\bf R}^n\times S^2$, where $S^2$
is a two-dimensional sphere of radius $r$.
The spherical coordinates of $S^2$ are denoted by $(\theta, \phi)$.
The model is consisted of a complex scalar field $f$ in 
${\bf R}\times S^2$ accompanied by a gackground gauge field $A$ in $S^2$, which
represents the Dirac-Wu-Yang monopole\cite{DWY-monopole}.
The vacuum configuration of the scalar field minimizes the
classical energy functional
\begin{eqnarray}
\label{energy-monopole}
E&=& \int d\theta d\phi \, r^2 \, \sin\theta 
\left[
     \frac{1}{r^2}
 \left|
     \frac{\partial f_{\pm}}{\partial \phi}
 \right|^2 \right.\nonumber\\
&&+ \frac{1}{r^2 \sin^2\theta}
 \left|
    \frac{\partial f_{\pm}}{\partial \phi}
    -iq(\pm 1- \cos\theta)f_{\pm}
 \right|^2\nonumber\\
&&
 -\mu^2 f^*_{\pm}f_{\pm} +\lambda (f^*_{\pm}f_{\pm})^2
\left.\frac{}{}\right].
\end{eqnarray}
In this section we examine the case of $q=\frac{3}{2}$, where three
vortices appear in $S^2$.\footnote{See ref.\cite{Vortex-J,DWY-monopole}
for more details.} 
The lowest expansion in a series of the eigenfunction is 
\begin{eqnarray}
\label{lowest-expansion}
f^{\frac{3}{2}}_{\pm}(\theta,\phi)&=&[
c_{3/2}e^{-3i\phi/2}\cos^3(\theta/2)\nonumber\\ 
&&+c_{1/2} 3^{1/2}e^{-i\phi/2}\cos^2(\theta/2)\sin(\theta/2)\nonumber\\
&&+c_{-1/2} 3^{1/2}e^{i\phi/2}\cos(\theta/2)\sin^2(\theta/2)\nonumber\\
&&+c_{-3/2} e^{3i\phi/2}\sin^3(\theta/2)
]e^{\pm3i\phi/2}.
\end{eqnarray}
From eq.(\ref{energy-monopole}) and eq.(\ref{lowest-expansion}), one can
obtain the energy functional.
The minimum is found at 
\begin{equation}
\label{vacuum-c}
(c_{3/2},c_{1/2},c_{-1/2},c_{-3/2})=(-v,0,0,v).
\end{equation}
From eq.(\ref{vacuum-c}) and eq.(\ref{lowest-expansion}), one can find
the location of the zero points of the scalar field $f$ are located at 
$\phi=0, \frac{2\pi}{3}, \frac{\pi}{3}$ on $\theta=\frac{\pi}{2}$,
which are determined up to reparametrization invariance.

Before arguing about string defect configuration, we should consider
cosmological evolution that makes string production possible.
Although cosmological mechanism for producing brane defects is still
unclear, we know at least one possible mechanism.
Assuming that the idea of thermal stabilization in ref.\cite{Thermal}
works in our case, three branes will glue together to a point during a
period of the Universe.
Below the critical temperature, the three branes begin to fall apart and
the symmetry is 
spontaneously broken.\footnote{We
should be 
careful about the above assumption.
In ref.\cite{Thermal}, supersymmetry is assumed in the bulk so that the
forces between branes are hierarchically suppressed.
In this case the time ($t_1$) when vortices(branes) are produced 
is much earlier than the time ($t_2$) when vortices begin to fall apart, 
which justifies the analysis in ref.\cite{Thermal}.
On the other hand, however, as we are not considering explicit supersymmetry
in our toy model, the condition $t_2 \gg t_1$ is not clear.
Thus in the above example, where supersymmetry is not included
explicitly, we have to assume that the formation of the three
vortices occurs at least simultaneously at a point in $S^2$.}.
The incidental string formed in this way will be related to the windings
of $\phi$.

Let us consider a straight string configuration along $z$-axis in
four-dimensional spacetime.
Here we introduce polar coordinates $r_s$ and $\phi_s$ in
four-dimensional spacetime.
We introduce a new variable $\alpha$ and rewrite the configuration
(\ref{vacuum-c}) as
\begin{equation}
\label{vacuum-c2}
(c_{3/2},c_{1/2},c_{-1/2},c_{-3/2})
=(-\sqrt{2}v\sin\frac{\alpha}{2},0,0,\sqrt{2}v
\cos\frac{\alpha}{2}).
\end{equation}
It is easy to see that the three vertices are settled at their vacuum
when $\alpha=\frac{\pi}{2}$.
For $\alpha=0$, three vertices coincide at $\theta=0$, which
corresponds to the core of the string.\footnote{Schematic pictures are
given in fig.3.}
It is straightforward to find explicit string configuration by using
$\alpha(r_s)$.
In our example, the core structure of the incidental string configuration 
is different from the primary brane defects, where ``brane'' must
be delocalized in the core.
If the $Z_3$ symmetry of the three vortices is an exact symmetry, one
will find alice string 
by identifying $\phi=\phi_s/3$. 
On the other hand, however, one should be careful about the original motivation
for constructing models with three vortices on $S^2$.
These models are interesting because it might reproduce the required
three generations of the standard model.
Because we know that the three generations in the standard model are not
equal, 
we think it should be natural to assume that the effective $Z_3$
symmetry is explicitly (but softly) broken in more phenomenological settings.
Even if the $Z_3$ symmetry is explicitly broken and the regular triangle
is warped in the
low-energy effective theory, one can construct string configuration by
identifying $\phi=\phi_s$.

Finally we will discuss why these incidental brane defects are visible for an
observer on a brane.
As the defects that we have discussed above are
formed by the parameters {\bf in the bulk}, the observation of
such defects is not a trivial issue.
Of course these defects are invisible if the physics on our brane is 
irrespective of the parameters that are shifted in the defect.
To see such defects, at least one of the parameters are required to be
detectable on our brane. 
For example, in the above example for a string, one might assume some
fields in the standard model are localized on each branes and their
interactions are suppressed by the distance between branes.
In this case, since the three branes coincide in the core of the defect,
observable enhancement of the interaction will appear in the
scatterings mediated by the string.

\section{Conclusions and Discussions}
\hspace*{\parindent}
In the models of brane construction, the isometry of a compactified
space might be 
broken by branes.
In four-dimensional effective Lagrangian, the breaking of an isometry is
 seen as the spontaneous breaking of the corresponding effective symmetry.
Then one might naturally 
expect that there are various kinds of defects that will be implemented
by the spontaneous symmetry breaking.
These defects are parametrized by brane positions, and will be
classified into two categories.
Primary brane defect is formed by a local fluctuation of the position of
a brane, which winds (or wraps) around a non-contractible circle of a
compactified space.
The structure of a primary brane defect is solely determined by the isometry of
the compactified space.
On the other hand, incidental brane defect is formed by the local
fluctuations of the relative brane positions among more than one brane.
We showed some explicit examples of incidental brane defects and
discussed differences between primary defects.

In this paper we did not consider time-dependent configurations, such as
Q-balls or instantons.
We think it is quite interesting to discuss the cosmological consequences of
``incidental'' brane Q-balls that might be produced after chaotic brane
inflation.

\section{Acknowledgment}
We wish to thank K.Shima for encouragement, and our colleagues in
Tokyo University for their kind hospitality.

\begin{figure}[htb]
\caption{Domain wall that interpolates between two degenerated vacua,
 $\phi>0$ and $\phi<0$, is shown.
Wall(brane) and anti-wall(anti-brane) located at each fixed point in the
 fifth dimension are
 exchanged when one passes through the center line in the upper picture.}  
\end{figure}

\begin{figure}[htb]
\caption{The area of a triangle in the left picture determines the Yukawa
coupling of the fields that live on each vertex.
Considering 1-loop correction, the triangle tends to shrink to a
 point, which is shown in the right picture. }
\end{figure}

\begin{figure}[htb]
\caption{Vortices are denoted by blobs. In the left picture, which
 corresponds to $r_s \rightarrow \infty$, vortices are placed at
 the equator of $S^2$, $\theta=\pi/2$. 
In the right picture, which corresponds to $r_s \rightarrow 0$,
all vortices are placed at the north pole, $\theta=0$.}
\end{figure}

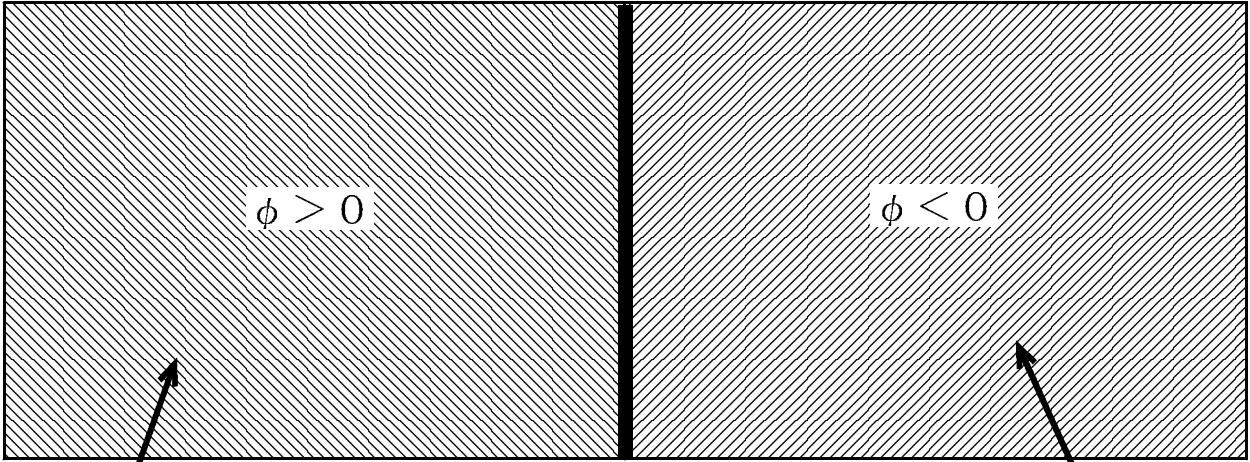
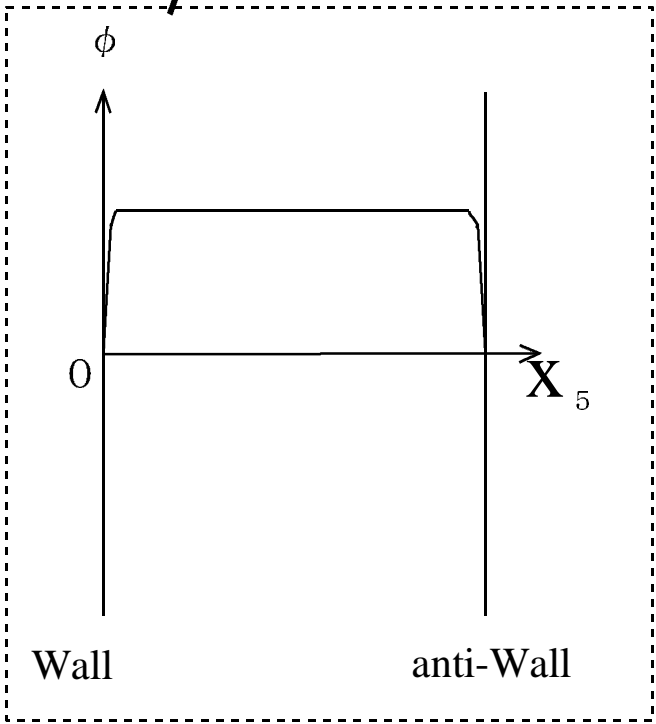
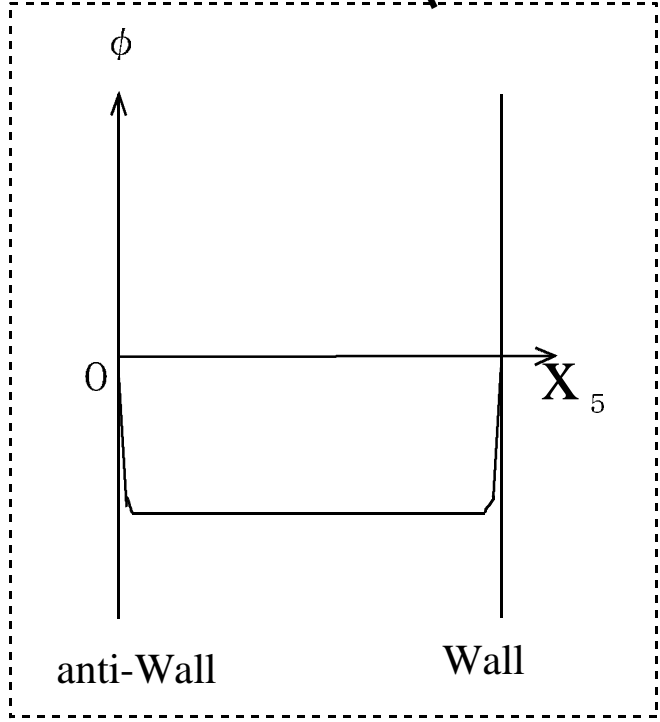
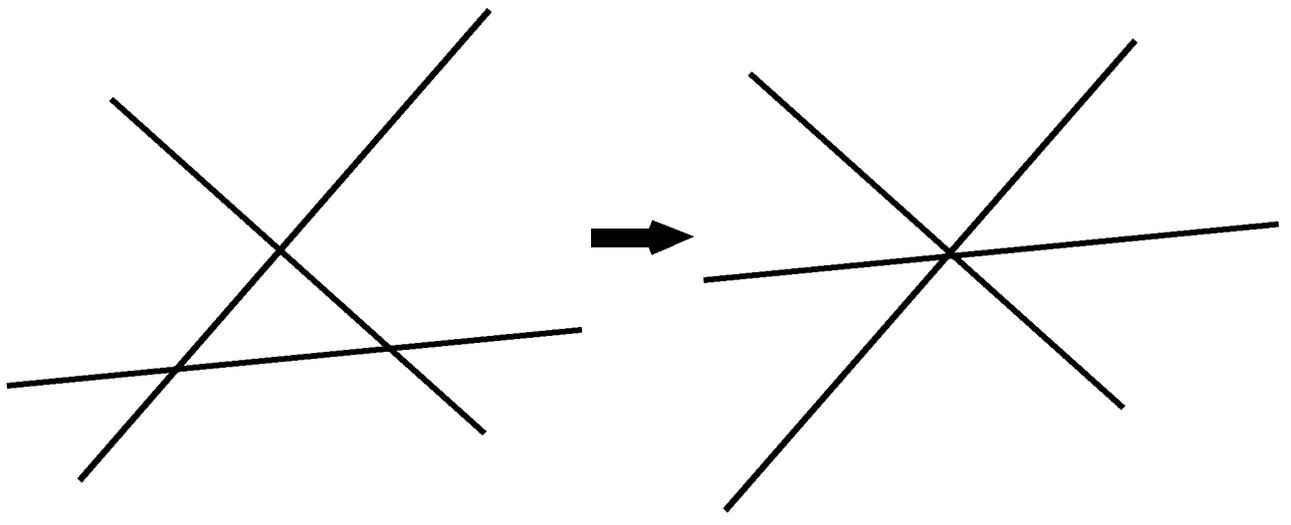
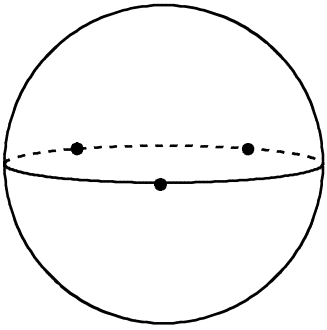
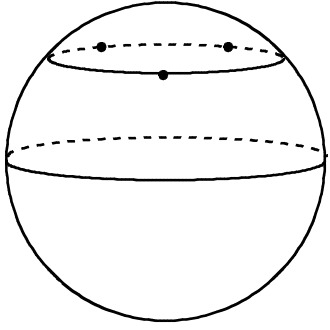
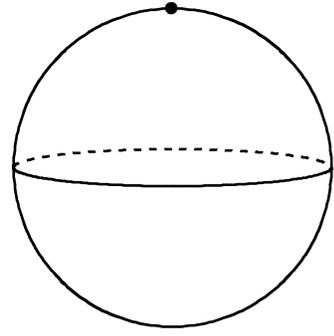
\end{document}